\documentclass[journal,comsoc,twoside]{IEEEtran}
\usepackage{graphicx}
\usepackage{multirow}
\usepackage{color}
\usepackage{soul}
\usepackage{mathrsfs}
\usepackage{subfig}
\usepackage{float}

\ifCLASSINFOpdf
\else
\fi

\begin{document}

\title{Enhancing RF Sensing with Deep Learning: A Layered Approach}

\author{Tianyue~Zheng,
	Zhe~Chen,
	Shuya~Ding,
	Jun~Luo,~\IEEEmembership{Senior~Member,~IEEE}%
}
\maketitle

\begin{abstract}

	In recent years, (Radio Frequency) RF sensing has gained increasing popularity due to its pervasiveness, low-cost, non-intrusiveness, and privacy preservation. However, realizing the promises of RF sensing is highly non-trivial, given typical challenges such as multipath and interference. One potential solution leverages deep learning to build direct mappings from RF domain to target domains, hence avoiding complex RF physical modeling. While earlier solutions exploit only simple feature extraction and classification modules, an emerging trend adds functional layers on top of elementary modules for more powerful generalizability and flexible applicability. To better understand this potential, this paper takes a layered approach to summarize RF sensing enabled by deep learning. Essentially, we present a four-layer framework: physical, backbone, generalization, and application. While this layered framework provides readers a systematic methodology for designing deep interpreted RF sensing, it also facilitates us to make improvement proposals and to hint future research opportunities.
\end{abstract}


\IEEEpeerreviewmaketitle

\section{Introduction}
Radio Frequency (RF) sensing is an emerging technology that is becoming increasingly active in the past decade. Compared with traditional device-based sensing, RF sensing has improved usability due to its contactless nature, along with other benefits such as pervasiveness, low-cost (in both money and energy), and privacy preservation. It can be used to estimate presence, status and activity of various targets, and has thus become the enabling technology for many high-level applications. Basically, RF signals (e.g., Wi-Fi, 4G/5G, Bluetooth) are ubiquitous in our surroundings. Targets such as humans in the transmission paths will reflect, refract, diffract, and scatter the RF signals following physical laws. By capturing these signals and processing the patterns, events in the environment can be sensed.

The goal of RF sensing is to ``translate'' the captured signal or its properties into meaningful results. Given certain efforts, Received Signal Strength Indicator (RSSI), Channel State Information (CSI), Time of Flight (ToF), Angle of Arrival/Angle of Departure (AoA/AoD), doppler frequency shift and even the baseband signal itself can be accessed at the physical layer. Traditional model-based RF sensing methods employ explicit mathematical models to describe the behavior of the RF signal propagation. Though widely adopted, these methods are inadequate in handling incomplete and inaccurate information, so their robustness is questionable.

One problem of model-based RF sensing is that the mathematical models oversimplify the real-life scenarios, and they tend to fail in complex environments. Factors including multipath fading and in-band interference can make the models extremely complicated and hence impractical. Moreover, model-based RF sensing cannot deal with high-level scene understanding tasks, such as human activity recognition (HAR) and pose estimation. Fortunately, deep learning offers us an alternative solution. As a data-driven approach, it enables us to extract RF features without explicitly modeling the underlying physical processes, thus removing the need for tuning the models manually. In addition, RF sensing based on deep learning has the potential to process high-dimensional data and to solve high-level problems.

Existing literature studies on the topic of RF sensing are abundant, but they focus mainly on Wi-Fi CSI (a special case of RF sensing), and lack an in-depth analysis of deep learning techniques~\cite{guo2017behavior,ma2019wifi}. In fact, recent developments have witnessed two major trends. On one hand, RF sensing exploiting signal sources beyond Wi-Fi has emerged. On the other hand, the application of deep learning is moving toward ``abstraction level'', instead of performing straightforward predictions in predefined settings. In particular, novel deep neural architectures proposed for RF sensing are able to distill cross-domain and environment-independent knowledge from the training data, leading to better generalizability.
Consequently, these trends call for a systematic and up-to-date survey on general RF sensing enabled specifically by deep learning.

\begin{figure}[ht]
	\centering
	\includegraphics[width=\linewidth]{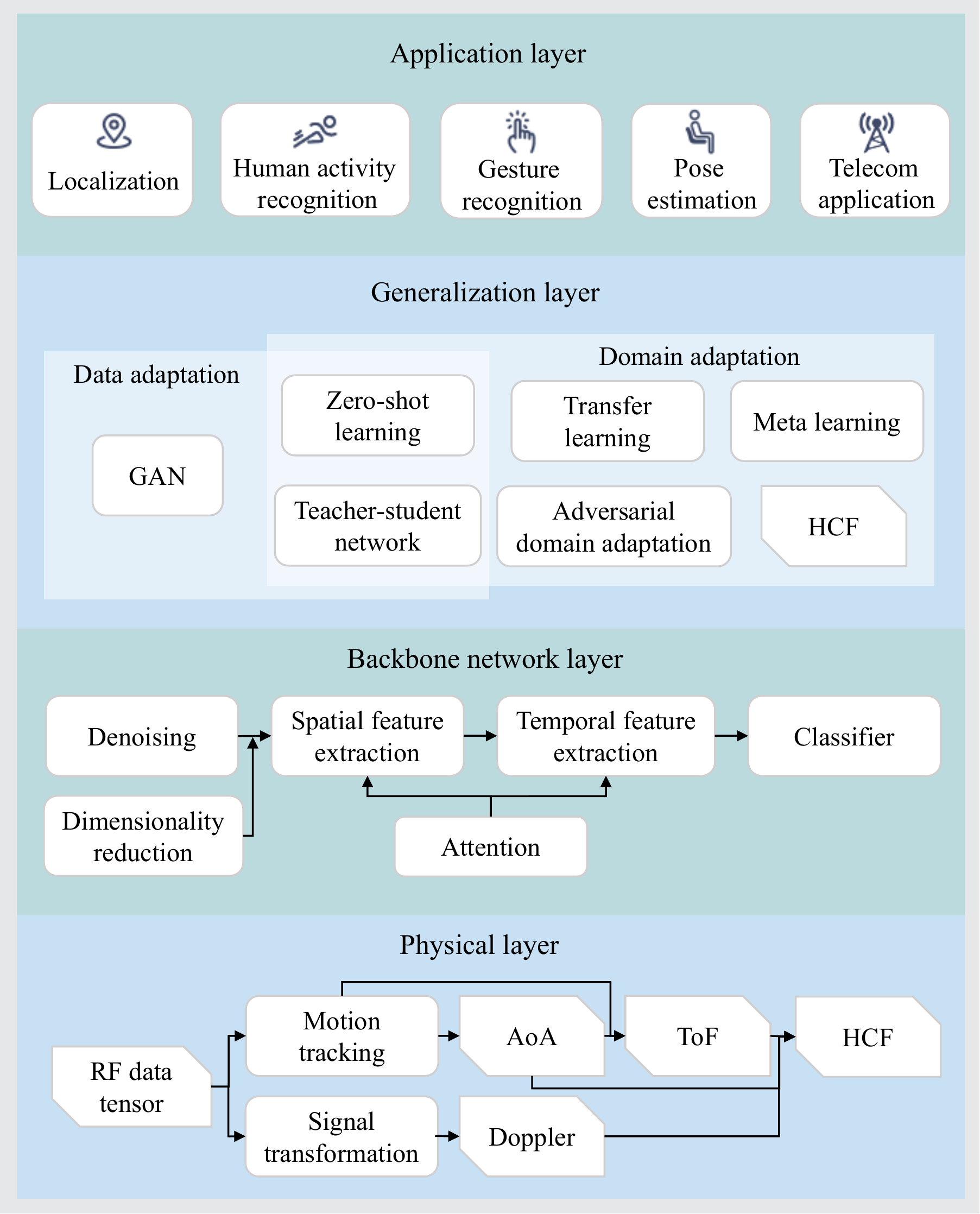}
	\caption{A layered framework for deep RF sensing.}
	\label{fig:arch}
\end{figure}

To this end, our survey takes a layered approach to capture the hierarchical designs for deep-learning-enabled RF sensing, aiming to provide a holistic and explainable framework on how these designs achieve higher-level generalization. Our framework encompasses four layers: physical, backbone, generalization, and application, as shown in Figure~\ref{fig:arch}. We first propose a unified data model for RF physical layer, bringing connected research directions into focus and laying the foundation of our framework. Then we review several major applications of RF sensing. Based on the physical model and application requirements, we survey the design of the deep learning architectures: i) the backbone layer for feature extraction and inference, and ii) the generalization layer for adapting to new domains and datasets. Improvement proposals are made for the backbone and generalization layers accordingly, along with possible future directions.

\section{Physical Layer}\label{physical}
The physical layer forms the bedrock of RF sensing, and is responsible for data collection and processing. In this section, we first introduce three physical schemes adopted by RF sensing, then we propose a model to unify the data delivered by different schemes.

\subsection{Schemes in the Physical Layer}
There exist three mainstream schemes for RF sensing:
\begin{itemize}
	\item{\textbf{Wi-Fi}:} Wi-Fi infrastructures can provide information for RF sensing without additional sensors~\cite{belmonte2019recurrent, jiang2018towards, zheng2019zero}. The information extracted from Wi-Fi are Received Signal Strength Indicator (RSSI) and Channel State Information (CSI). Whereas RSSI only provides coarse-grained signal intensity information~\cite{belmonte2019recurrent}, CSI is a fine-grained characterization of an Orthogonal Frequency-Division Multiplexing (OFDM) channel~\cite{jiang2018towards, zheng2019zero}, depicting the channel state of each subcarrier between every transmitter-receiver antenna pair. CSI can be obtained from  Wi-Fi cards with customized drivers. 

	\item{\textbf{Frequency-Modulated Continuous Wave}:} FMCW is a type of radar dedicated for sensing purpose~\cite{guan2020through, zhao2017learning}. It transmits a chirp signal (a signal whose frequency increases or decreases linearly in time). Because the frequency difference between the transmitted and reflected signals is proportional to the Time of Flight (ToF), the range of a target can be calculated by measuring this difference. Other information can be further obtained, for example, speed can be estimated by the Doppler frequency offset.  

	\item{\textbf{Impulse-Radio Ultra-WideBand}:} IR-UWB~\cite{fhager2019pulsed} is also a type of radar dedicated for sensing, but with a different waveform design. Different from FMCW's chirp signals, IR-UWB transmits pulses with a very short duration, thus occupying a large bandwidth. Consequently, the range of a target can be directly obtained by measuring the time delay of the reflected signals, which in turn helps to deduce other information. 
\end{itemize}

Both Wi-Fi and FMCW employ an antenna array to exploit multiple-input and multiple-output (MIMO) advantages, so MIMO is assumed as default setting hereafter. Also, as millimeter waves (mmWave) are adopted by Wi-Fi (partially as IEEE 802.11ad) and FMCW radar, it is well covered by our following discussions. 

\subsection{A Unified Data Model of the Physical Layer} \label{ssec:cube}
\begin{figure}[ht]
	\centering
	\includegraphics[width=.88\linewidth]{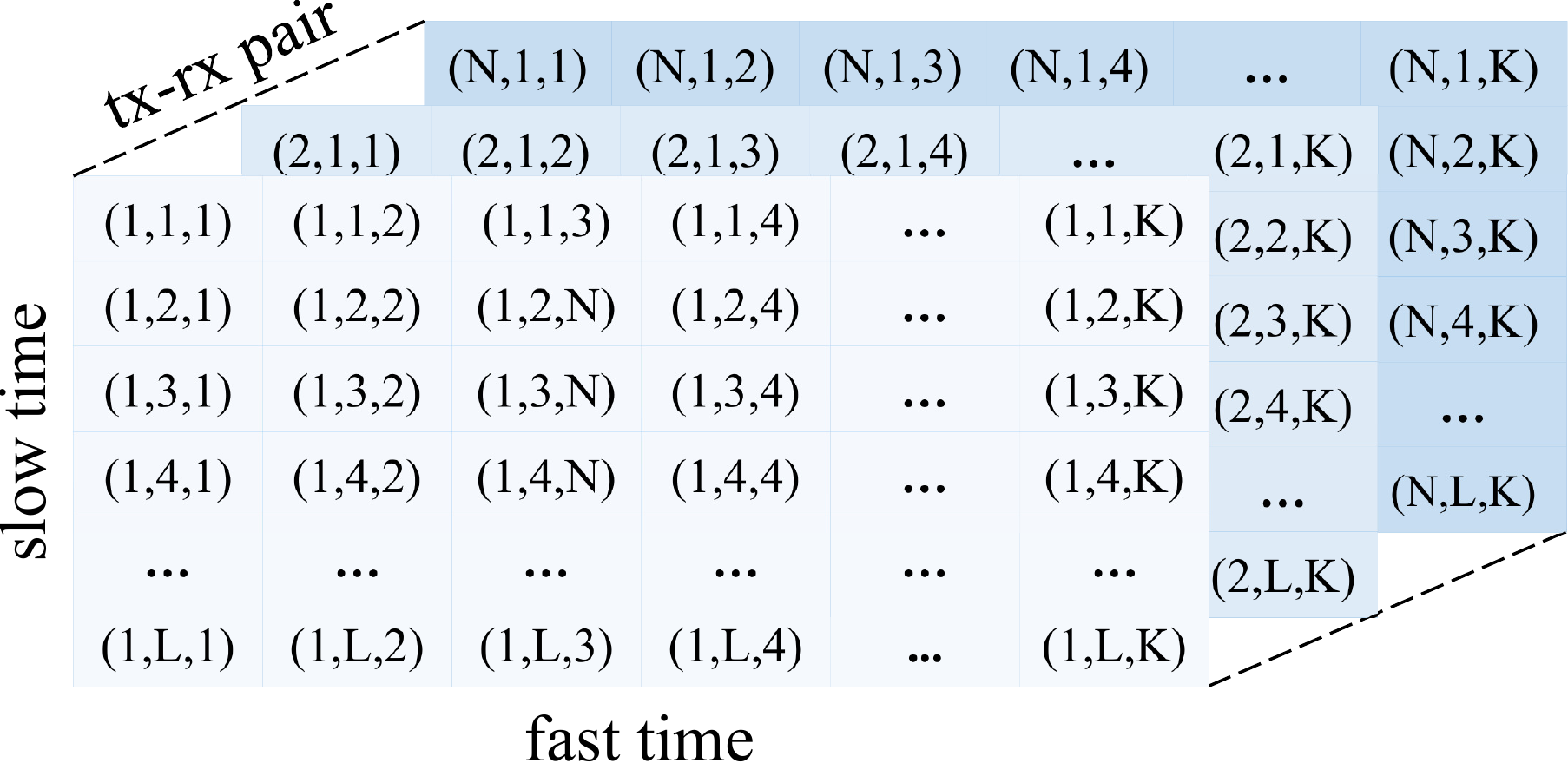}
	\caption{A unified data model for the physical layer.}
	\label{fig:physical}
\end{figure}

In order to aid upper deep learning layers in processing RF data from heterogeneous schemes, we hereby propose a data model to unify Wi-Fi, FMCW, and IR-UWB. Essentially, we represent the RF data by a complex tensor $\mathbb{C}^{N \times K \times L}$ with three dimensions: fast-time, slow-time, and tx-rx antenna pair (i.e., a pair of transmitting-receiving antenna elements in the respective MIMO antenna arrays of sender and receiver), as shown in Figure~\ref{fig:physical}. The fast-time dimension (length $K$) represents the samples (potentially transformed via FFT) of a Wi-Fi OFDM symbol, FMCW chirp, or IR-UWB pulse. The slow-time dimension (length $L$) describes how the symbol/chirp/pulse repeat and change in the long run. The tx-rx dimension (length $N$) describes the spatial diversity.

Based on the unified model, Fourier transform (or wavelet transform) can be further performed along the slow-time axis to obtain Doppler characteristics, which in turn suggests the velocity of a moving target. Spatial spectrum analysis algorithm (e.g., Capon, MUSIC, and ESPRIT) can be applied to the tx-rx dimension to estimate the target direction~\cite{ma2019wifi}. This unified RF data representation founds the basis for further processing by deep learning layers. 

\section{Application Layer}\label{application}
The development of RF sensing has enabled many interesting IoT applications. Here we focus only on the most important applications (listed in Table~\ref{comparison2}), although occupancy sensing, user identification, material classification also belong to the ecosystem of the application layer. We choose to introduce this layer before the deep learning layers as the neural network designs are driven by applications. 

\begin{table*}[ht]
	\centering
	\caption{A comparison of recent works on RF sensing with deep learning.}
	\label{comparison2}
	\begin{tabular}{|l|l|l|l|l|}
		\hline
		Paper                                        & Application        & Backbone Network           & Generalization Technique         \\ \hline
        Belmonte et al. 2019~\cite{belmonte2019recurrent}& Localization       & RNN             & GAN         \\ \hline  %
        Li et al. 2019~\cite{li2019af}               & Localization       & CNN                    & GAN                              \\ \hline   %
        Jiang et al. 2018~\cite{jiang2018towards}    & HAR                & CNN                    & Adversarial domain adaptation    \\ \hline   %
        Ding et al. 2020~\cite{ding2020rfnet} & HAR & CNN+LSTM & Meta learning \\ \hline
        Zhao et al. 2017~\cite{zhao2017learning}     & Sleep monitoring               & CNN+RNN                & Adversarial domain adaptation    \\ \hline   %
        Zheng et al. 2019~\cite{zheng2019zero}       & Gesture recognition               & CNN+RNN              & HCF                         \\ \hline   %
        Fhager et al. 2019~\cite{fhager2019pulsed}     & Gesture recognition        & CNN               & Transfer learning\\ \hline   %
        Tamzeed et al. 2019~\cite{tamzeed2019wi}     & Gesture recognition              & CNN+LSTM               & Zero-shot learning               \\ \hline   %
        Yang et al. 2020~\cite{yang2020mobileda}     & Gesture recognition               & CNN+LSTM               & Teacher student network          \\ \hline   %
        Wang et al.  2019~\cite{wang2019person}      & Pose estimation    & U-Net+attention         &  Adversarial domain adaptation                \\ \hline   %
        Guan et al. 2020~\cite{guan2020through}     & Imaging                & CNN               & GAN             \\ \hline   %
        Yang et al. 2019~\cite{yang2019deep} & Channel prediction & FNN & Transfer learning \\ \hline
	\end{tabular}
\end{table*}

\subsection{Localization}
The growing need for location-based services call for accurate localization. RF sensing provides a solution of localization in indoor environments where GPS fails to reach. Furthermore, data-driven approaches are increasingly employed with RF sensing for localization purposes~\cite{alrabeiah19}. As an example, \textit{RF fingerprinting} matches measured features (e.g., distilled from RSSI and CSI) to predefined ones on an RF map. However, this method requires lots of fingerprints to cover the space. Also, the performance is affected by the high variability of RF signals caused by multipath fading and interference. Therefore, deep learning comes into play: by fully exploring RF data and extracting deep features, irrelevant information can be stripped while useful spatial features retained~\cite{belmonte2019recurrent, li2019af}, and localization accuracy can be potentially improved.

\subsection{Human Activity Recognition (HAR)}
RF-based methods rival computer vision in HAR applications, since they are robust to low-light conditions, work in the existence of occlusions, and preserve privacy. In addition, they achieve good recognition accuracy while maintaining low processing and deployment costs. Deeming reflected RF signals at any moment as a \textit{snapshot} of human status, a time series of such snapshots may indicate certain human activity. Therefore, an RF-HAR task is equivalent to classifying a time series of RF snapshots. Deep learning can be employed to extract the spatial-temporal features from the RF data, then a detected activity will be chosen from a class of trained human activities. Recent RF-HAR applications include, among others, motion recognition~\cite{jiang2018towards}, gesture recognition~\cite{zheng2019zero, tamzeed2019wi, yang2020mobileda, fhager2019pulsed} and sleep stage detection~\cite{zhao2017learning}. 

\subsection{Pose Estimation}
Pose estimation is the process of locating human joints, limbs, and torso. More specifically, RF signals reflected off human body are analyzed, so as to extract the skeletons or the meshes of human body. However, this is not an easy task, since RF signals do not carry the coordinates of human body parts. To make things worse, there are many joints and limbs in the human body, and different parts are associated with different poses. As a result, the degree of freedom of pose estimation is much higher than that of a simple localization or classification task, rendering traditional model-based methods unable to handle this ill-posed problem. In this context, deep learning, together with prior knowledge of the human body~\cite{jiang2018towards, wang2019person}, has been used to recover human poses from the incomplete information contained in RF signals.  %

\subsection{Other Applications}
A few applications beyond the above categories are also worth mentioning. \textit{Imaging} employs sensing devices to perceive the target and create a two-dimensional image at radio wavelengths. In~\cite{guan2020through}, FMCW radar is used to reconstruct a high-resolution image of the objects behind dense fog, assisted by deep learning. \textit{Channel estimation} is a core challenge in wireless communications, aiming to counteract the adverse effects of multipath, shadowing, and mobility. As the non-linearity of deep learning network makes it suitable for handling non-linear distortion, interference and frequency selectivity caused by a channel, it is adopted to predict downlink CSI and to recover the transmitted signal~\cite{yang2019deep}.

\section{Backbone Network Layer}\label{base}
With the RF data provided by the physical layer and the application-driven goal in mind, we are ready to discuss the design of deep learning layers. We focus on introducing the building blocks of a backbone network (illustrated in Figure~\ref{fig:arch}) in this section. We also propose our idea on how to assemble the modules together for RF sensing tasks at the end.

\subsection{Data Preprocessing}
Before feeding the RF data into neural networks, some preprocessing is needed. One of them is denoising, which removes noise and irrelevant information from an environment. Common approaches for denoising include filtering and Principal Component Analysis (PCA)~\cite{ma2019wifi, tamzeed2019wi}. Filtering means separating signal and noise by transforming the RF signals into another domain. PCA, instead, projects signals onto uncorrelated ``principal components'', so that the components with the largest variances are kept (as a larger variance indicates more information); others are deemed noise and discarded. Furthermore, PCA pre-processes data via dimensionality reduction, which may lead to a more robust model that tends not to overfit the data.

\subsection{Spatial Feature Extraction}\label{ssec:spatial}
The RF data is ready for spatial feature extraction after preprocessing. The spatial features are extracted from the 2D slices consisting of the tx-rx pair and the fast-time dimension, as mentioned in Section~\ref{ssec:cube}. While different pairs of tx-rx antennas contain angular information of the targets, the fast-time dimension contains range information of the targets. Traditionally, model-based approaches directly compute angular and range information of a target and exploit them to further derive spatial features. However, deep learning networks often skip these intermediate steps and extract spatial features in an end-to-end fashion.

\begin{itemize}
	\item{\textbf{Feedforward Neural Network (FNN)}:} As the most rudimentary form of deep learning, FNN is universal in the sense that is capable of approximating arbitrary functions and thus extracting arbitrary spatial features. However, redundant parameters of fully connected FNN makes it inefficient to train and tend to overfit, therefore only a limited number of existing proposals~\cite{yang2019deep} use FNN for feature extraction in localization and HAR.

	\item{\textbf{Convolutional Neural Network (CNN)}:} To overcome the drawbacks of FNN, CNN encodes the prior knowledge of translation invariance into itself. On one hand, the convolution operation gives a constant activation value regardless of the feature location; on the other hand, the pooling layer downsamples the feature map to make it insensitive to the change of feature location. As such, CNN takes orders of magnitude less training data than FNN, therefore it has become the widely adopted spatial feature extractor~\cite{jiang2018towards,zheng2019zero, zhao2017learning,  tamzeed2019wi}. 

\end{itemize}

As a side note, the features mentioned in this section are ``spatial'' in the sense that they capture structural information on the 2D slice of the RF tensor, instead of providing direct information of the 3D space (e.g., AoA and range). For now, FNN and CNN as fitting functions cannot provide this information reliably. We will discuss how to improve this in Section.~\ref{ssec:spi} in order to develop deep learning models with better interpretability and explainability. 

\subsection{Temporal Feature Extraction} \label{ssec:temporal}
The spatial features introduced in Section~\ref{ssec:spatial} is enough for differentiating static objects. However, for tasks like HAR in which targets do not stand still, temporal contextual information is needed to capture the target dynamics.  The temporal features can be extracted along the slow-time axis from the RF tensors, as mentioned in Section~\ref{ssec:cube}. We now introduce several temporal feature extraction modules.
\begin{itemize}

	\item{\textbf{Recurrent Neural Network (RNN)}:} The first category of temporal feature extractor is RNN. It not only takes the current value as input but also considers past inputs, which is stored as hidden states of the RNN nodes. The nodes form a directed graph closely related to the time sequence, hence displaying temporal behavior. Compared to CNN, RNN is able to process input series of any length, and the complexity of the model does not increase with the input size, therefore RNN is chosen in~\cite{belmonte2019recurrent, zhao2017learning} for extracting temporal features.

	\item{\textbf{Long Short-Term Memory Network (LSTM)}:} Another temporal feature extractor is LSTM. It is similar to RNN but can maintain information in memory for longer periods. LSTM solves the vanishing and exploding gradient problem of long RNN by introducing \textit{forget gate}, \textit{input gate}, and \textit{output gate} into the network. The gates work together to decide how much information will be thrown away, kept in the node, taken as input and used as output. By controlling the flow of information, LSTM outperforms RNN when dealing with long dependencies, as demonstrated in~\cite{jiang2018towards, tamzeed2019wi, yang2020mobileda}. 

\end{itemize}

\subsection{Attention Mechanism}

Attention mechanism is a weighted aggregation method for optimizing deep learning models. It can be understood as ``focusing'' on more important segments of the RF tensor, i.e., important parts are assigned with larger weights and therefore having greater influence on the result. The attention module works as a trainable add-on of the feature extractor~\cite{wang2019person}. For spatial RF features, the attention mechanism together with CNN allows prioritizing areas where the target is located. For temporal features, the attention mechanism can be used with RNN/LSTM to focus on the segments where important events occur. It is shown that the attention mechanism improves the exploration of the relationship between the input RF data and the inference result, hence enhancing the performance of the deep learning model for RF sensing~\cite{wang2019person}.

\subsection{Summary and Potential Improvements}\label{ssec:spi}
Equipped with all the building blocks, we propose the following pipeline to summarize the backbone network layer for feature extraction and inference. First, RF tensors (as mentioned in Section~\ref{ssec:cube}) are pre-processed to achieve dimensionality reduction. Second, the data are fed to both spatial and temporal feature extractors, with the latter focusing on sequence learning. Both feature extractors are assisted by an attention module for better focusing on useful data. At last, the output features are used for pattern recognition or classification. In fact, several backbone networks can be juxtaposed to process multiple input streams (including the original RF tensor and transformed tensor in other domains), for fully exploiting the high-dimensional data. Whereas such a pipeline seems to have become a de facto standard, there are still research opportunities within this layer:

\begin{itemize}

	\item Complex-valued neural network is a potential improvement of the backbone layer. As mentioned in Section~\ref{ssec:cube}, elements of the input tensor are complex numbers, therefore an accommodating complex-valued network with a richer representational capacity may be desirable.
    \item CNN has shown to be too popularly employed in Section~\ref{ssec:spatial}, but it is computationally expensive, and does not encode orientation and relative position information. Nonetheless, AoA and range estimation are essentials for MIMO array processing. Therefore, we suggest to explore other possibilities (e.g. the capsule network) for constructing the backbone network layer.
	\item Existing backbones are mostly designed for classification. However, other applications may adopt different designs; these include: %
	R-CNN for object detection, U-Net for semantic segmentation, and CNN+RNN+attention for RF question answering. Essentially, the vast body of literature and experience from both computer vision and natural language processing need to be explored to innovate the design of the backbone layer.

\end{itemize}

\section{Generalization Layer}\label{generalization}

Traditionally, a single backbone network in Section~\ref{base} is enough for deep learning tasks.  However, this architecture faces two challenges: \textit{domain shift} and \textit{data scarcity}. \textit{Domain shift} indicates that the performance drops when domain (environment, user, or device) changes; \textit{data scarcity} means that not enough training data is collected. Meanwhile, the two are closely related: domain shift entails a laborious process of collecting and on-line labeling new data. Previously, researchers in~\cite{zheng2019zero} try to solve the challenges by handcrafting domain-independent features (HCF). However, HCFs depend on prior information such as position and orientation, hence incurring another laborious process in obtaining this information and making the efforts futile.

\begin{figure*}[t]
	\setlength\abovecaptionskip{5pt}
	\setlength\belowcaptionskip{-1pt}
	\centering
	\includegraphics[width=\linewidth]{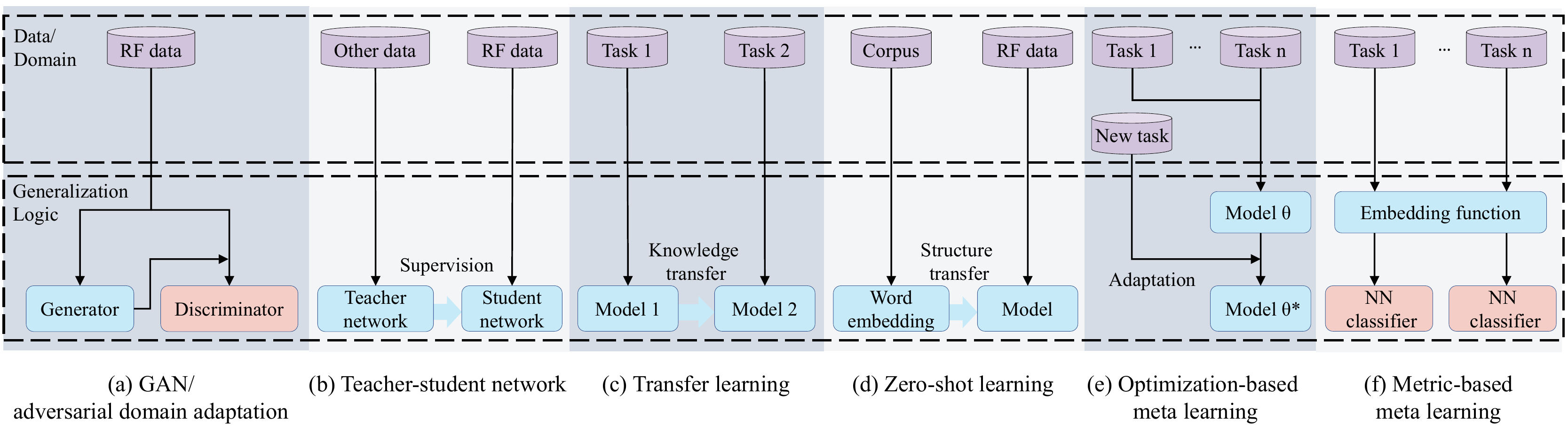}
	\caption{Different techniques used in the generalization layer.}
	\label{fig:generalization}
\end{figure*}

As a solution, we propose a generalization layer on top of the backbone network, as shown in Figure~\ref{fig:arch}. It extends the capability of the backbone which is only trained for a single domain. Figure~\ref{fig:generalization} illustrates the functionality of the generalization layer with different techniques: by employing corresponding generalization logic (e.g. knowledge transfer and data generation), this layer generalizes the backbone model from the training domain to the ``unseen'' target domains. As a result, deep learning models for RF sensing system can be trained once, but adaptive to other places, people, or devices, hence providing more extensible services. Next, we summarize previous generalization techniques, and also put forward our proposal.

\subsection{Generative Adversarial Network}\label{ssec:gan}

Generative Adversarial Network (GAN) in Figure~\ref{fig:generalization}(a) helps generalize from small training datasets to larger, more universal datasets. GAN is a two-player game, in which the generative network ($\mathcal{G}$) and the discriminative network ($\mathcal{D}$) are pitted against each other. During training, $\mathcal{G}$ generates fake data conforming to the distribution of a given dataset, trying to confuse $\mathcal{D}$; while $\mathcal{D}$ the classifier, tries to discriminate fake data generated by $\mathcal{G}$ and true training data. After a few rounds, the data generated by $\mathcal{G}$ is so good so as to fool $\mathcal{D}$, which then can be used to extend the dataset. In~\cite{li2019af}, Li et al. use $\mathcal{G}$ to generate Wi-Fi fingerprints for indoor localization. By extending the dataset, the RF sensing system achieves a faster convergence time and requires less human effort. Belmonte et al.~\cite{belmonte2019recurrent} employ GAN to recover lost fingerprints for continuous localization and tracking.

\subsection{Adversarial Domain Adaptation}
Adversarial domain adaptation in Figure~\ref{fig:generalization}(a) is an approach to generalize from one domain to other domains. It aims to learn models that transform different domains into a common feature space. It is closely related to GAN mentioned in Section.\ref{ssec:gan}, where there is a competition between $\mathcal{G}$ and $\mathcal{D}$. The difference is that in GAN, the output of $\mathcal{G}$ is of interest, whereas in adversarial domain adaptation, we are interested in learning features that are invariant across domains by playing the min-max game. In~\cite{jiang2018towards}, adversarial domain adaptation is used to achieve environment-independent HAR. Wang et al.~\cite{wang2019person} employ this method to transform and adapt the RF data tensor to new environments for pose estimation. Zhao et al.~\cite{zhao2017learning} applies it to make learning human sleeping stages independent from individuals and measurement conditions.

\subsection{Teacher Student Network}
Teacher student network in Figure~\ref{fig:generalization}(b) is another candidate for generalization layer. The network is traditionally used for knowledge distillation: a larger teacher network $\mathcal{T}$ with many parameters ``teaches'' a much smaller student network $\mathcal{S}$ how to perform tasks. $\mathcal{S}$ is trained by copying the exact behavior of $\mathcal{T}$. Since $\mathcal{T}$ is wider and deeper, it tends to have better generalizability. By letting $\mathcal{T}$ guiding $\mathcal{S}$, a similar generalizability can be achieved by $\mathcal{S}$. The teacher-student structure is mainly used when the application imposes limitations on a solution (e.g., insufficient training data and limited computation power), therefore it must resort to a larger network for more powerful generalization. In MobileDA~\cite{yang2020mobileda}, the $\mathcal{S}$ network is deployed on edge device, therefore its training is supervised by the $\mathcal{T}$ network on the server for more generalized knowledge. 
\subsection{Transfer Learning}\label{ssec:transfer}

Transfer learning in Figure~\ref{fig:generalization}(c) works by fine-tuning pre-trained model to handle similar tasks, since these tasks often share common knowledge. For example, the trained weights of the first several layers of neural networks can be reused across different tasks, since the layers capture input features independent of specific tasks. We can transfer this knowledge and fine-tune the rest of the model, therefore adapt the model to new task with few samples and low cross-domain training cost. In our generalization layer, the idea is used to transfer knowledge from the source domain to the target domain (e.g., RF environments, datasets, devices, and human subjects). Fhager et al.~\cite{fhager2019pulsed} employs transfer learning to achieve accurate cross-domain gesture classification. Yang et al.~\cite{yang2019deep} leverage the generalizability of transfer learning to predict the CSI of FDD MIMO downlink in previously unseen environments.

\subsection{Zero-Shot Learning}

Zero-shot learning in Figure~\ref{fig:generalization}(d) intends to recognize classes whose instances are unseen during training. Instead of relying on lots of training data, it recognizes new classes by descriptions. More precisely, zero-shot learning exploits knowledge from semantic space, and then relate it to known classes, hence generalizing to more classes. For example, in HAR, in order to teach the concept of a \textit{run} to the model that has already learned \textit{walk}, instead of training the networks with many RF data of people running, zero-shot learning enables the network to recognize \textit{run} by incorporating the relationship between \textit{run} and \textit{walk} (learned from the semantic space) into its previous knowledge of \textit{walk}. In Wi-Fringe~\cite{tamzeed2019wi}, researchers transfer the knowledge from attribute embeddings of English words to RF sensing systems for recognizing unseen human activities.

\subsection{Optimization-Based Meta Learning}
Meta learning, which literally means ``learning to learn'', is the state-of-the-art technique to improve the generalizability of deep neural networks. Optimization-based meta learning in Figure~\ref{fig:generalization}(e), which is a type of meta learning, achieves fast domain adaptation by tweaking the gradient descent algorithm. Model-Agnostic Meta Learning (MAML) is an implementation of optimization-based meta learning. It initializes the model with parameters that can be adapted to any new task using a few gradient descents. MAML has been proven to be applicable to a variety of different tasks, including classification, regression, and reinforcement learning. This generalization technique has not been applied to RF sensing yet, possibly due to its high complexity that confines the depth of backbone networks.

\subsection{Our Proposal}
\begin{figure}[h]

	\centering
	\subfloat[Number of training samples.]{
	\begin{minipage}[b]{0.7\linewidth}
		\centering
		\includegraphics[width=.9\linewidth]{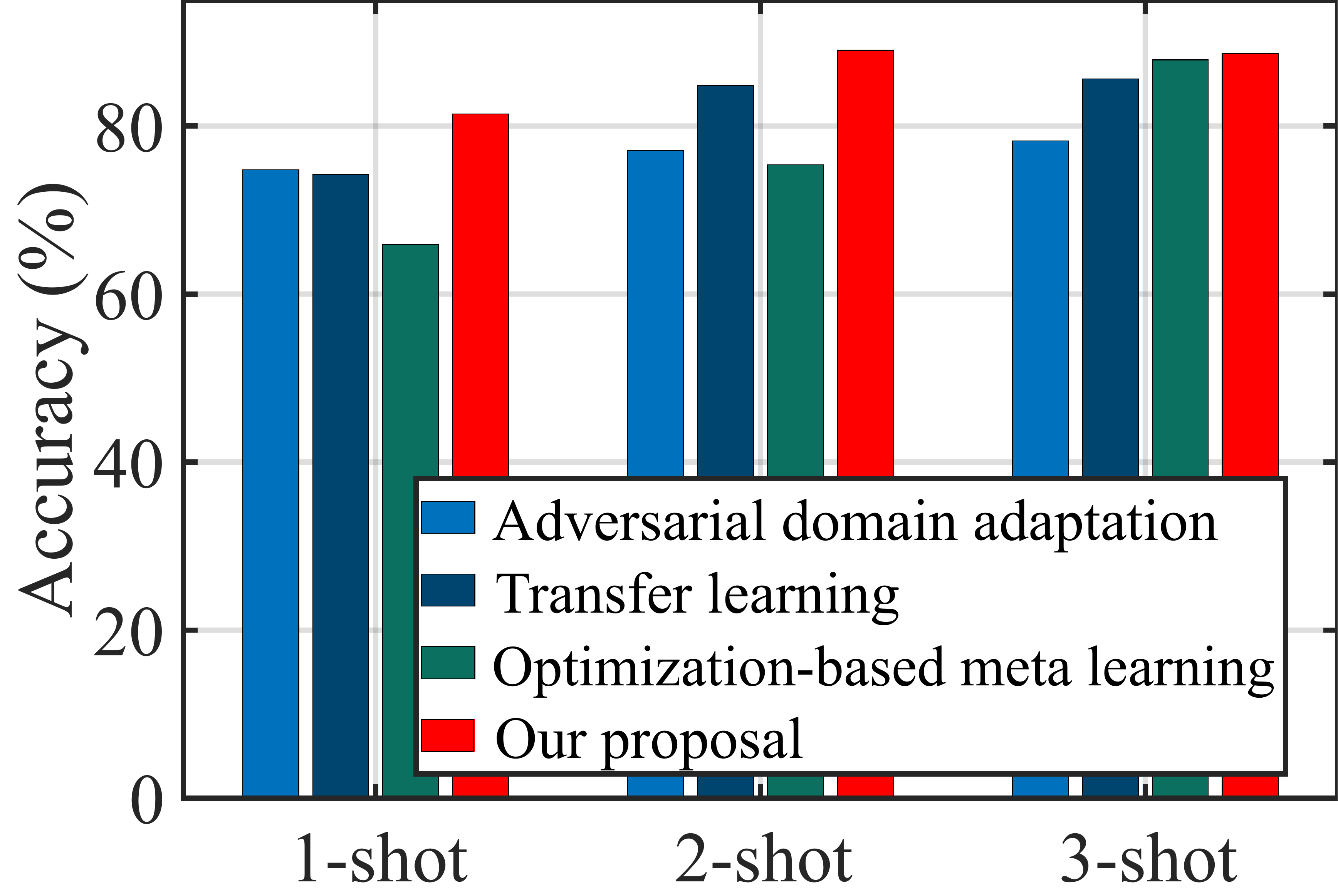}
		\label{fig:fmcw_accuracy}
	\end{minipage}
}
\\
	\subfloat[RF technologies.]{
		\begin{minipage}[b]{0.7\linewidth}
			\centering
			\includegraphics[width=.9\linewidth]{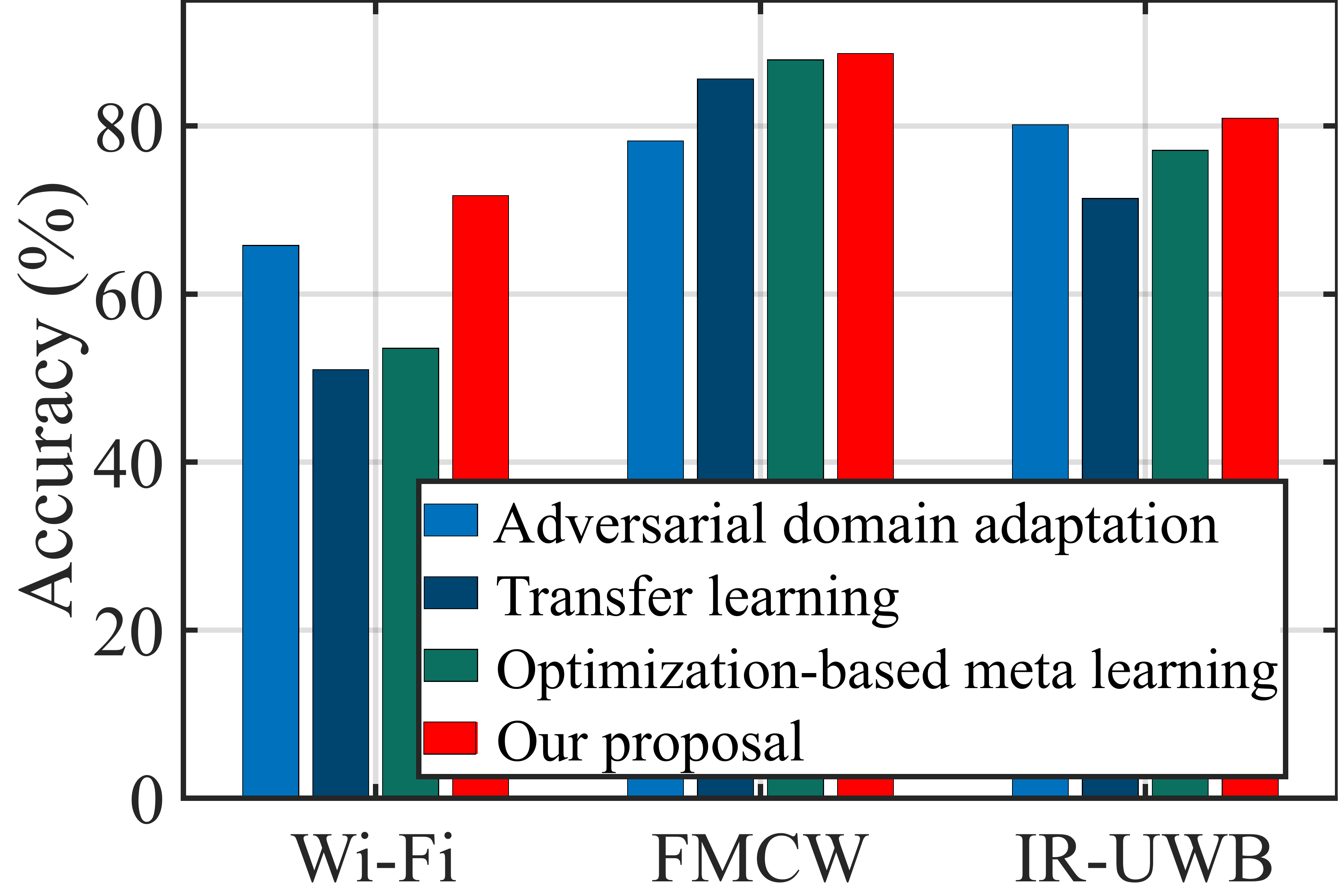}
			\label{fig:wireless_accuracy}
		\end{minipage}
	}
\caption{Performance of four typical generalization techniques under different experiment conditions.}
\label{fig:trajectory}
\vspace{-1ex}
\end{figure}

The generalization layer is still open for researchers to investigate new strategies. As optimization-based meta learning is slow in training, adds system complexity, and fails to handle large backbone networks and unbalanced classes, we hereby propose metric-based meta learning~\cite{ding2020rfnet} as an alternative. As shown in Figure~\ref{fig:generalization}(f), metric-based meta learning is simple, yet flexible and powerful. Instead of learning a classifier, this method tries to learn a good \textit{embedding function}, which encodes RF data to a feature vector. By measuring the similarity (via Euclidean norm, cosine distance, or a learned metric) among feature vectors, RF tensors can be classified to its nearest neighbor (NN). The embedding function can be quickly re-trained in new domains, hence expediting new learning tasks. Preliminary HAR experiment results involving Wi-Fi, FMCW, and IR-UWB are used to demonstrate the effectiveness of our method. The center frequencies of these three devices are 5.8, 77, and 7.29~\!GHz, respectively; and their respective bandwidths are 0.04, 4, and 1.4~\!GHz. They are tested in 100 new domains, i.e., 10 rooms, each with 10 distinct layouts and placements of objects. Figure~\ref{fig:fmcw_accuracy} shows the recognition accuracies of 10 human subjects performing 6 activities for different generalization techniques. Under only one, two, and three labeled training samples for each class (namely ``1-shot", ``2-shot", and ``3-shot") in a new domain, our proposed model exhibits a superiority over other methods. Similarly, our method consistently outperforms others for all physical-layer schemes in Figure~\ref{fig:wireless_accuracy}.

In addition to this brief proposal, we further highlight a few potential research directions:

\begin{itemize}
    \item Self-supervised learning utilizes the intrinsic relation of unlabeled data for supervision. It can teach the neural network some ``common sense'' in the RF data, which is useful for generalizing to target domain where high-quality labels are hardly available.
	\item Network architecture search (NAS) is a technique for automating the backbone network design, aiming to select the best-performing architecture after traversing various choices. One may design end-to-end learning pipelines with better generalizability by employing NAS.
	\item Incorporating prior knowledge of the underlying physical process may also help. Instead of developing pure model-based techniques, priors derived from the (physical) domain shift process can potentially bring both new insights and better explainability. 

\end{itemize}

\section{Conclusion}\label{conclusion}
In this article, we take a layered approach to summarize deep-learning based RF sensing. The layered framework includes four layers, namely, physical, backbone, generalization, and application. In the physical layer, a novel RF data model provides a standard data interface for upper layers. Based on the unified physical layer representation and driven by various applications, we discuss the design methodologies of the backbone network layer that converts complicated RF sensing modeling to end-to-end deep learning tasks. We also add a generalization layer on top for more powerful generalizability and flexible applicability. This article provides a comprehensive summary of this emerging field, presents a proper theoretical framework and thus motivates researchers to design new deep-learning enabled RF sensing systems.  

\section{Acknowledgements}
This research is supported in part by AcRF Tier 2 Grant MOE2016-T2-2-022 and AcRF Tier 1 Grant RG17/19.

\vskip 0.4in

\noindent \textbf{Tianyue Zheng} (tianyue002@e.ntu.edu.sg) received his B.Eng. degree in Telecommunication Engineering from Harbin Institute of Technology, China, and M.Eng. degree in Computer Engineering from the University of Toronto, Canada. He is currently working towards his Ph.D. degree in Computer Science from Nanyang Technological University, Singapore. His research interests include RF sensing and deep learning.

\vskip 0.2in

\noindent \textbf{Zhe Chen} (chen.zhe@ntu.edu.sg) is a research fellow in Nanyang Technological University, Singapore. He received the Ph.D. degree with honor in Computer Science from Fudan University, Shanghai, China, and obtained Doctoral Dissertation Award from ACM SIGCOMM China 2019. His research interests include RF communication and sensing systems, deep learning, and IoT applications.

\vskip 0.2in

\noindent \textbf{Shuya Ding} (di0002ya@e.ntu.edu.sg) received her B.Eng. degree in Electrical and Electronic Engineering from Nanyang Technological University (NTU), Singapore. She is currently working towards her Ph.D. degree in Computer Science from NTU. Her research interests include machine learning and its practical applications.

\vskip 0.2in

\noindent \textbf{Jun Luo} (junluo@ntu.edu.sg)  received his BS and MS degrees in Electrical Engineering from Tsinghua University, China, and the Ph.D. degree in Computer Science from EPFL, Lausanne, Switzerland. He is currently an Associate Professor at School of Computer Engineering, Nanyang Technological University. His research interests include mobile and ubiquitous computing and applied operations research.

\end{document}